\documentclass[conference]{IEEEtran}
\usepackage[utf8]{inputenc}
\usepackage{amsmath}
\usepackage{amssymb}
\usepackage{graphicx}
\usepackage{url}
\usepackage{xcolor}

\usepackage[]{fancyhdr} %
\newcommand{\changefont}{\fontsize{7}{7}\selectfont}
\newtheorem{remark}{Remark}{}
\newtheorem{lemma}{Lemma}{}
{}
\newtheorem{thm}{Theorem}{}
\newtheorem{definition}{Definition}{}

\IEEEoverridecommandlockouts
\title{Cyber-Resilient Frequency Control of Power Grids with Energy Storage Systems}
\author{\IEEEauthorblockN{Jairo Giraldo and Masood Parvania}
\IEEEauthorblockA{Department of Electrical and Computer Engineering, The University of Utah, Salt Lake City, UT, USA\\
Email: {\{jairo.giraldo, masood.parvania\}@utah.edu }}}

\begin{document}

\maketitle
\thispagestyle{fancy}
\pagestyle{fancy}

\begin{abstract}
The integration of synchronous generators and energy storage systems  operated through communication networks introduces new challenges and vulnerabilities to the electric grid, where cyber attacks can corrupt sensor measurements or control inputs and interrupt functions such as frequency regulation.
This paper proposes a defense methodology for the design of resilient operating constraints imposed on each generation and storage unit in order to prevent any attack sequence from driving the system's frequency to unsafe conditions. 
The resilient operating constraints are found by using   ellipsoidal approximations of the reachable set of the power system, leading to a convex optimization problem with linear matrix inequalities. 
Numerical results in a single-area power system with synchronous generation and energy storage demonstrate how the resilient constraints provide security guarantees against any type of attack affecting frequency measurements or controller setpoints.
    
\end{abstract}

\vspace{5pt}

\section{Introduction}
The integration and coordination of synchronous generators and energy storage systems (ESS) in power grids have the potential to increase the reliability and stability of the grid.
However, to fully harness the potential benefits of ESS, it is necessary to integrate monitoring and communication technologies that can make the power system vulnerable to cyber attacks that aim to disrupt its operation  \cite{li2012securing}. These vulnerabilities  have become more evident in the last couple of years where several cyber attacks have affected power systems. For instance, a synchronized and coordinated cyber attack compromised three Ukrainian regional electric power distribution companies, resulting in power outages affecting approximately 225,000 customers for several hours \cite{case2016analysis}. Similarly, a denial-of-service attack  left grid operators temporarily blinded to generation sites of several dozens of wind and solar farms in the U.S. \cite{SECURITY}.

A significant amount of effort has been made to address different cyber-security challenges in power systems. For instance, authors in \cite{Ameli2018} utilize multiple unknown input observers to detect any discrepancy in frequency deviations and identify the system parameters that were corrupted. Ganjkhani et al., \cite{ganjkhani2021} present a novel anomaly detection and localization strategy that uses deep neural networks to differentiate between faults and cyber attacks, and simultaneously find the location of the affected line or device in the power distribution system. 
Similarly, in \cite{khan2022}, a real-time reference model that mimics the nonlinear and complex behavior of power distribution networks is used to detect false-data injection attacks through the computation of residuals and a chi-square detection score. 
On the other hand, to minimize the social and economical costs caused by cyber attacks in power systems, the grid must be not only secure to prevent or detect intrusions, but also resilient to \emph{withstand successful attacks} \cite{clark2017cyber}. Cyber resilience of the power grid refers to the capability of maintaining critical functionality of the grid in the presence of cyber attacks \cite{arghandeh2016definition,nguyen2020electric}. Few works have proposed mechanisms that enhance the resilience of the power grid in the face of cyber threats. 
In \cite{chen2019resilient}, authors introduce a resilient power sharing mechanism that is able to cancel the influence of attacks 
 affecting the controller computation by using state observers to estimate the attacker's action. Similarly, authors  in \cite{khalaf2018joint} utilize input and state estimators to detect and compensate the effects of cyber attacks in sensor measurements intended for the computation of the automatic generation control (AGC) setpoints.  Diaz-Garcia et al., \cite{diaz2021resilient} propose a methodology to redesign the communication network of distributed control in the power grid in order to reduce the impact caused by false-data injection attacks. Authors in \cite{giraldo2022MTD_TSG} introduce a decentralized moving-target defense methodology that is able to reduce the impact of cyber attacks by replicating relevant signals and selecting them randomly.
 
 Three major limitations are identified from previous work. 
First,  anomaly detection and mitigation are not sufficient to guarantee system security, because it is not possible to design algorithms that can detect all types of attacks.  Second, most of the contributions found in the literature are tailored  for very specific types of attacks that lack the generality needed in real applications where it is hard to make assumptions about the attacker's actions. Finally, the existent solutions to enhance resiliency focus mostly on  modifying a control algorithm or the communication network architecture to make the system less susceptible to certain attacks. However, they do not offer any guarantee that the system can remain operating in safe conditions for a wide range of attack scenarios.  

In this paper, the aforementioned limitations are addressed in order to  guarantee  the resilience of power systems affected by cyber attacks. To this end, in this work we  propose a novel methodology   that uses the dynamic model of the power grid to find optimal resilient operating constraints for generation units and energy storage systems that ensure that no attack affecting the AGC setpoints directly (e.g., tampering communication between the control center and local controllers) or indirectly (e.g., compromising frequency measurements used to compute the area control error (ACE)) can drive the system to unsafe states, i.e., those states that threaten system stability and may lead to equipment damage or even large-scale blackouts.   One of the main benefits of the proposed approach is that it is agnostic to the type of controller, the type of attack, and does not require any prior assumption about the attacker's capabilities. The exact computation of these constraints becomes an  NP-hard combinatorial problem when the  range of operation of the generation and storage units and the system states are  assumed to be  continuous (i.e., bounded infinite sets); however, we overcome these issues by using ellipsoidal approximations of the reachable set, simplifying the complexity of the optimization problem and  allowing the use of linear matrix inequalities (LMI) that can be efficiently solved using   interior-point methods for semidefinite programming.  

The remainder of this paper is organized as follows: Section \ref{sec:system_description}  introduces the power grid dynamic model. Section \ref{sec:resilient} presents the proposed optimization problem   to find the resilient operating constraints of each synchronous generator and energy storage system. The numerical results are presented in Section \ref{sec:case_study}, and the conclusions are drawn in Section \ref{sec:conclusions}.

\section{System Description}\label{sec:system_description}
\subsection{Dynamic Model of the Power Grid}
\begin{figure}
    \centering
    \includegraphics[width=\columnwidth]{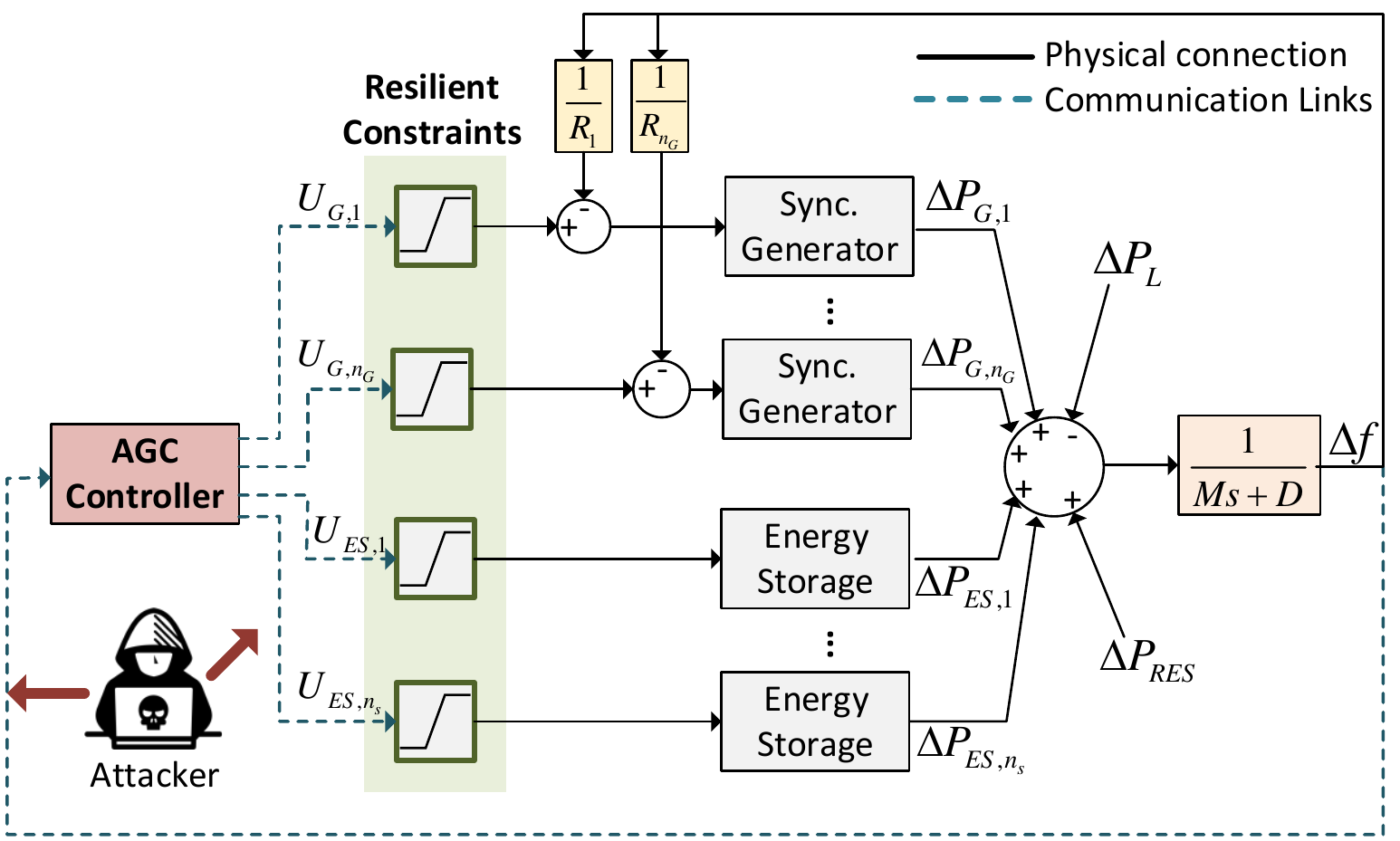}
    \caption{Block diagram representation of a single-area power system with synchronous generators, energy storage system, and primary and secondary frequency control. A cyber attacker can tamper frequency measurements used for the AGC setpoint computation or directly AGC setpoints sent to the different generation and storage units. }
    \label{fig:general}
\end{figure}

Consider the system in Fig. \ref{fig:general}, which consists of  a set of  dispatchable  synchronous generators $\mathcal{N}_G=\{1,\ldots, n_G\}$ that combine bulk generation and small generators (e.g., diesel generators, small-hydro power plants), and a set of energy storage systems $\mathcal{N}_{ES}=\{1,\ldots, n_{ES}\}$, where each one may represent individual or aggregated energy storage units. The AGC manages generation units and energy storage in order to maintain the frequency deviations within safe limits after disturbances caused by loads and renewable energy resources. 
Let $\Delta P_G=[\Delta P_{G,1},\ldots,\Delta P_{G,n_G}]^\top$,
denote the deviation of the generated power with respect to a scheduled setpoint for each dispatchable synchronous generator. Similarly, let $\Delta P_{ES}=[\Delta P_{ES,1},\ldots, \Delta P_{ES,n_{ES}}]$ denote the deviation of the charging/discharging scheduled setpoint for each energy storage unit. The change in the aggregated  load is denoted as $\Delta P_{L}$. Renewable energy resources (e.g., solar/wind generators) can contribute to supply part of the demand, such that the change of the aggregated renewable resources is denoted as  $\Delta P_{RES}$. Both, $\Delta P_L,\Delta P_{RES}$ are unknown uncertainties for the system. The control setpoints computed by the AGC are  $U_{G}=[U_{G,1},\ldots, U_{G,n_{G}}]^\top$ and $U_{ES}=[U_{ES,1},\ldots U_{ES,n_{ES}}]$.  
The frequency deviation in the system is then dictated by the aggregated inertia  and damping coefficients $M$ and $D$, respectively, and is described as follows
 \begin{equation}\label{eq:freq}
     \dot \Delta f(t)=-\frac{D}{M}\Delta f(t)+\frac{1}{M}\Delta P_{T}(t), 
 \end{equation}
where \begin{align*}\Delta P_T(t)=&\sum\nolimits_{j=1}^{n_G}{\Delta P_{G,j}(t)}+\sum\nolimits_{l=1}^{n_{ES}}{\Delta P_{ES,l}(t)}\\
&+\Delta P_{RES}(t)-\Delta P_{L}(t).\end{align*}   

The  active power deviation  of each synchronous generator $j\in\mathcal{N}_G$ can be modeled using a first order dynamic equation of the form \cite{Wang2018,Bevrani2009}:
\small
\begin{align}\label{eq:gen}
    \dot \Delta P_{G,j}(t)&=-\frac{1}{T_{t,j}}\Delta  P_{G,j}(t)+\frac{1}{T_{t,j}}\Delta X_{gov,j}(t)\nonumber\\
    \dot \Delta X_{gov,j}(t)&=-\frac{1}{T_{g,j}}\Delta  X_{gov,j}(t)-\frac{1}{T_{g,j}R_j}\Delta f(t)+\frac{1}{T_{g,j}}U_{G,j}(t),
\end{align}
\normalsize
where $\Delta X_{gov}=[\Delta X_{gov,1},\ldots,\Delta X_{gov,n_G}]^\top$ denotes the change in the governor position, $T_{t,j}$ is the turbine time constant, $T_{g,j}$ is governor time constant, $R_j$ is speed regulation gain from the primary frequency control. Even though our results can be easily extended for other generation models  such as hydropower plants \cite{doolla2011load} or steam units \cite{Bevrani2009}, for simplicity we focus on synchronous generators that can be described by \eqref{eq:gen}.  
The deviation of the absorbed/injected power provided by the energy storage systems can be described as follows:
\begin{equation}
    \Delta P_{ES,i}=-\frac{1}{T_{ES,i}}\Delta P_{ES,i}+\frac{1}{T_{ES,i}}U_{ES,i},
\end{equation}
where $T_{ES,i}$ is a time constant that represents the transient between the received setpoint and the actual injected/absorbed power. For most storage systems $T_{ES,i}$ is small due the capability of ESS to quickly deliver/absorb power. 

Let $x\in\mathbb{R}^n$ with $x=[\Delta f,\Delta P_G^\top,\Delta X_{gov}^\top,\Delta P_{ES}]^\top$ denote the system states,     $u\in\mathbb{R}^{m}$, with $u=[U_G^\top,U_{ES}^\top]^\top$ denotes the control input, and $\omega\in\mathbb{R}$ with $\omega=\Delta P_{L}-\Delta P_{RES}$ corresponds to the external unknown disturbance. The combined dynamics can be described in a compact form as follows:
\begin{equation}\label{eq:xdot}
    \dot x(t)=A_cx(t)+B_cu(t)+H_c\omega(t),
\end{equation}
with matrices $A_C,B_c,H$ of proper dimension.
Given that the AGC setpoints are typically computed every $\tau$ seconds (e.g., typically 2-5\;s), in the remaining of this work we will focus on discrete-time dynamics. However, the simulation results will be tested on a continuous-time model with piece-wise continuous control commands updated every $\tau$ seconds. 

The discrete-time representation of \eqref{eq:xdot} is then
\begin{equation}\label{eq:x}
    x(k+1)=Ax(k)+Bu(k)+H\omega(k),
\end{equation}
where $A,B,H$ are computed as $A=e^{\tau A_c }$ and $B=\int\nolimits_{0}^{\tau}{e^{A_c\tau}B_cd\tau}$, $H=\int\nolimits_{0}^{\tau}{e^{A_c\tau}H_c d\tau}$. 
\subsection{Adversary Model}
In this paper, we consider a powerful adversary that knows the system dynamics (i.e., matrices $A,B,H$),  can compromise any type of measurements and/or  all AGC setpoints,   and  has the ability to change them arbitrarily. The attacker is not constrained to follow specific attacks such as replay, delay, scaling or bias attacks previously considered in the literature, but instead can inject any type of attack sequence. The attacker's goal is to force the system to operate in unsafe conditions that may potentially damage the system equipment or cause a large-scale blackout. 
These strong assumptions will allow us to design a defense mechanism that is able to withstand any type of attack, either on sensor measurements or AGC setpoints, which will lead to provable security guarantees.

\section{Resilient Operating Constraints}\label{sec:resilient}
The power system dynamics described above depend on the AGC setpoints, $u(k)$, and the unknown external disturbance $\omega(k)$. The execution of each control input $u_i(k)$ is bounded according to physical constraints specific to each type of generator or energy storage, such that  $|u_i(k)|\leq \gamma_i$, for $\gamma_i>0$ and $i=1,\ldots, m$. For instance, a diesel generator can generate up to a limited amount of power depending on its physical characteristics; a battery can absorb/inject power with maximum and minimum power ratings depending on its construction. Similarly, load deviations caused by changes in the electricity consumption and the power supplied by RES are bounded according to physical limits, such that $|\omega(k)|\leq \gamma_{\omega}$. 
Despite these physical constraints, it is possible for cyber attackers to design attack sequences that drive the system states to unsafe operation conditions (e.g., frequencies deviations above/below 0.2 Hz). In fact, the integration of ESS may increase the attacker's surface making the power grid more susceptible to cyber attack impacts. For instance, if an attacker that manipulates an ESS  sends charging commands during high electricity demand, it may cause power imbalances that lead to dangerous frequency deviations. Therefore, in  this work, new operating constraints imposed on each generator and  ESS will be designed   (see Fig. \ref{fig:general}) to guarantee that the power grid can withstand any attack sequence, either in sensor measurements or control commands. To this end, ideas related to the ``reachable set'' will be adopted, which have been widely studied for defining the stability of dynamic systems, and then a convex optimization problem will be formulated to find the optimal resilient operating constraints.


\subsection{Reachable Set Approximation}
The reachable set with respect to the system states $x$ can be defined as the set of all admissible $x$ that satisfy the physical constraints,  described as follows:  
\small
\begin{equation}\label{eq:feasible_set}
    \mathcal{R}:=\left\{x(k)\in \mathbb{R}^n \left|\begin{array}{l}
    x(k)=Ax(k)+Bu(k)+H\omega(k)\\
    | u_i(k)|\leq \gamma_i,\; i=1,\ldots,m+1,\\
    |\omega(k)|\leq \gamma_\omega,\;\forall k\in\mathbb{N}\end{array}\right.\right\}.
\end{equation}
\normalsize
Note that the reachable set contains all the states that can be reached by any combination of input sequences $u(0),u(1),\ldots,u(k),\ldots $ and $\omega(0),\omega(1),\ldots,\omega(k),\ldots$. Therefore, the reachable set provides valuable information about any system states that can be attained by potential attack sequences that AGC setpoints.

It is possible to estimate  the reachable set $\mathcal{R}$ by finding points that satisfy the equality and inequality constraints. However, since each $u_i$ and $\omega$ can take values in a continuous space, $\mathcal{R}$ is an infinite set, and computing the exact reachable set (or at least a very good approximation) is difficult and  computationally costly when the number of variables increases. For this reason, we will exploit some properties from control theory and convex optimization  to find the smallest ellipsoid $\mathcal{E}(W)$ that contains the reachable set, i.e., $\mathcal{R}\subseteq\mathcal{E}(W)$. 

\begin{figure}
    \centering
    \includegraphics[width=\columnwidth]{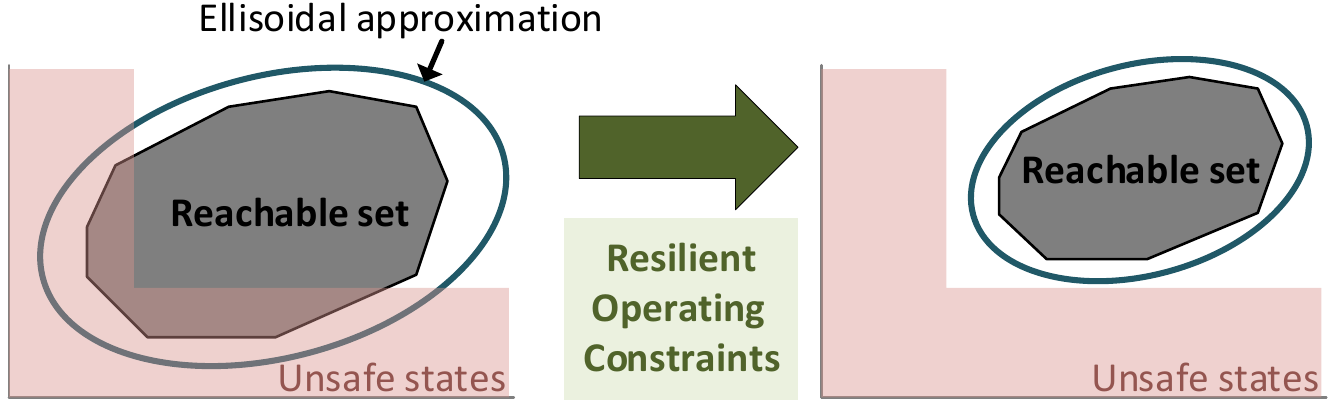}
    \caption{Illustration of the proposed methodology using the reachable set and  its ellipsoidal approximation. The resilient operating constraints are designed to guarantee that unsafe states are not reachable by any potentially malicious control input.}
    \label{fig:summary}
\end{figure}

Typically, an ellipsoid of dimension $n$ can be described in a quadratic form as follows:
\begin{equation}
\mathcal{E}(W):=\left\{ x \in \mathbb{R}^n\ |\ x^T W^{-1}x\leq 1 \right\},
\end{equation}
where $W \in \mathbb{R}^{n \times n}$ is a positive-definite matrix. Figure \ref{fig:summary} (left) illustrates an example of the ellipsoidal approximation that contains the reachable set. Notice that the reachable set may overlap unsafe states. These unsafe states represent the set of system states that are dangerous and must be avoided, e.g., $\Delta f(t)\geq 0.2\;Hz$ or $\Delta f(t)\leq -0.2\; Hz$. 
In other words, if the reachable set overlaps with the unsafe states, it implies that there exist attack sequences that can drive the power grid to unsafe states. 
Therefore, the main goal of this work is to find new resilient constraints  for each generator and ESS such that the reachable set does not overlap the unsafe states, as illustrated in Fig. \ref{fig:summary} (right).  



\subsection{Finding the Resilient Operating Constraints}

 Using tools from control theory and convex optimization we will expand the  methodology introduced in \cite{giraldo2020daria,kafash2018constraining} to approximate the entire reachable set with an ellipsoid $\mathcal{E}(W)$. Then, a new group of input constraints will be designed  to ``shrink'' the reachable set in a way that dangerous states are no longer reachable, as illustrated in Fig. \ref{fig:summary}. As a consequence, it would be guaranteed that any attack sequence affecting sensor measurements or AGC setpoints cannot drive the system states to dangerous states.  These new resilient constraints may be enforced by local controllers at each generation unit and ESS, such that any AGC setpoint outside these bounds will be saturated, making the system resilient to any cyber attack affecting the AGC  directly or indirectly. First, let us introduce a mathematical formulation for the unsafe states set.

\begin{definition}[Unsafe States]
 The  set of unsafe states, denoted as  $\mathcal{D}$,  corresponds to those states that are dangerous for the system operation and that need to be avoided. This set can be captured as  the union of half-spaces defined by  hyperplanes of the form $c_i^\top x\geq g_i$, such that 
 \begin{equation}\label{eq:unsafe}
\mathcal{D}=\left\{x\in\mathbb{R}^N\left|\bigcup\limits_{j=1}^\kappa{c_j^\top x\geq g_i}\right.\right\},\end{equation}
\end{definition}
 As an example, consider $\Delta f(k)$. Recall that the safety condition for frequency deviation is $|\Delta f(k)|\leq 0.2\;Hz$. In this case, the unsafe states can be described by two hyperplanes with parameters $c_1=[1,0,\ldots,0]$, $g_1=0.2$, and $c_2=[-1,0,\ldots,0]$ and $g_2=0.2$. The same formulation applies for all the states of $x$. 
 
The shortest distance between an ellipsoid $\mathcal{E}(W)$ and each hyperplane that characterize the unsafe states, $c_i^\top x\geq g_i$ is given by \cite{kurzhanskiy2006ellipsoidal} 
 \begin{equation}\label{eq:distance}
d_{\mathcal{E},i}=\frac{|g_i|-\sqrt{c_i^\top Wc_i}}{c^\top c_i}.\end{equation}
Therefore, since our intent is to redesign the  operating constraints in order to avoid dangerous states, we can formulate an  optimization problem  to find new constraints $\hat \gamma_i$ such that the distance $d_\mathcal{E,j}$ between the ellipsoid and each hyperplane is greater than zero for all $j=1,\ldots,\kappa$. This in turn will guarantee that the unsafe states $\mathcal{D}$ and the new reachable set $\mathcal{\hat  R}$ do not overlap. The main result is introduced in the following theorem. 

\begin{thm}[{Resilient Operating Constraints Design}] \label{thm:dangerous}
Consider the linear time-invariant (LTI) system \eqref{eq:x} with controllable pair $(A,B)$ and a set of dangerous states $\mathcal{D}$ defined by \eqref{eq:unsafe}. If $A$ is Hurwitz stable, and if there exists an $a \in{(0,1)}$  for which the positive definite matrix $W$ is a solution of the following convex optimization problem:\\
\noindent
\begin{equation} \label{thmdangerous}
\left\{\begin{aligned}
	&\min_{W,\hat \gamma_1,\ldots,\hat\gamma_m}\ -\sum\limits_{l=1}^{m}{\hat\gamma_l},\\
    &\text{s.t.}\ W>0,\ \hat\gamma_i\leq \gamma_i,\ \text{and}\\
    &\quad \begin{bmatrix}
		aW & 0 & 0& WA^\top\\
		 0 & \frac{(1-a)}{m}\hat R_u & \hat R_uB^\top \\
		 0&0&\frac{(1-a)}{m}\gamma_\omega^2 &\gamma_\omega^2H^\top\\
		 AW & B\hat R_u & H\gamma_\omega^2 &W
	\end{bmatrix} \geq 0, \\
    &\quad c_i^\top W c_i\leq  g_i^2,\qquad i=1,...,\kappa,
\end{aligned}\right.
\end{equation}
then the new operating bounds $\hat \gamma_i$, $i=1,...,m$, ensure that the resulting reachable set $\mathcal{\hat R}$ does not intersect with the dangerous states $\mathcal{D}$, as depicted in Fig. \ref{fig:summary}(right). 
\end{thm}
\begin{IEEEproof}[Sketch of the proof]
The proof consist of guaranteeing that a positive definite function $V(k)=x(k)^\top W^{-1}x(k)$ remains bounded for all $k$, along the trajectories of the dynamic system. This can be ensured if the following is satisfied:
\[V({k+1})-aV(k)-\frac{(1-a)}{m}u(k)^\top R_u^{-1}u(k)-\frac{\omega(k)^2}{\gamma_{\omega}^2}\leq 0,
\vspace{-5pt}.\]
In addition, the condition $d_{\mathcal{E},i}\leq 0$ is equivalent to $c_i^\top W c_i\leq  g_i^2$. Finally, the objective function is set to maximize  the resilient operating constraints, such that  frequency regulation is still maintained. 
The details of the proof are  found in Appendix~\ref{app:thm1}.
\end{IEEEproof}
\begin{remark}
Theorem \ref{thm:dangerous} contains an unknown parameter $a$ that enters nonlinearly with the variable $W$. To side-step the nonlinearity that is caused by the products $aW$, it is possible to perform a grid-search or bisection over the parameter ``$a$,'' such that the optimization is repeated a number of times for different values of $a$. The selected solution $\hat R_u$ corresponds to the one with the largest $\sum\nolimits_{l=1}^{m}{\hat \gamma_l}$.
\end{remark}

\textcolor{black}{
The solution of Theorem \ref{thm:dangerous} maximizes each $\hat \gamma_i$ to ensure that the system without attack  can still perform frequency regulation.  Since the optimization problem in \eqref{thmdangerous} is convex and solved offline, it is possible to find resilient operating constraints even for large numbers of controllable assets. 
Even though the proposed solution does not affect the system stability and guarantees safety, it could be possible that the obtained constraints decrease the frequency regulation performance (e.g., longer times to reach $\Delta f=0$) when compared to the nominal case without resilient constraints, incurring in additional operating costs. This may require the use of incentives from system operators (e.g., transmission system operator) so that the obtained resilient constraints are enforced by local controllers of energy assets participating in frequency regulation.}

 \section{Case Study}\label{sec:case_study}
 In order to illustrate the viability of the proposed approach, we consider a power system with a single area, two synchronous generators, and two energy storage systems. It is assumed that at $t=0$ the frequency deviation starts at $\Delta f(0)=0.1\;Hz$. The total external input corresponding to the uncertainty caused by load changes and  renewable energy resources is unknown but bounded according to $|\Delta P_{L}-\Delta P_{RES}|\leq 0.2\;pu$. Every $\tau=2$ seconds, the controller receives the frequency measurements to compute the area control error given by $ACE(k)=-\mathcal{B}\Delta f(k)$, such that 
 \[AGC(k)=K_PACE(k)+K_I \sum\nolimits_{0}^{k}ACE(k),
 \] 
 where $\mathcal{B}=10,K_P=0.1,K_I=10$. 
 Therefore, the AGC setpoints transmitted to each generator and energy storage unit is given by 
 \[u_i(k)=\beta_i AGC(k),\]
 where $\beta_i$ is the participation factor of each unit, for $\sum\nolimits_{i=1}^m{\beta_i}=1$ \cite{wang2022}. For the simulations, we consider $\beta=[0.3, 0.4, 0.2, 0.1]$.
 The parameters of the generators and ESS are summarized in Table \ref{table:parameters} 
  \begin{table}[h]
 \caption{Power system parameters}
\label{table:parameters}
\begin{tabular}{l|l}

\hline
\multicolumn{2}{c}{Parameters Synchronous Generators} \\ \hline
Approx. total inertia constant  $M$    & 5 pu$\cdot$s/Hz \\
Approx. total damping constant  $D$    & 3 pu/Hz \\
Governor time constants ($T_{g,1},T_{g,2}$)   & (0.8, 0.12)\;s  \\
Turbine time constants ($T_{t,1},T_{t,2}$)    & (3, 0.5)\;s     \\
Energy storage time constants ($T_{ES,1},T_{ES,2}$) & (0.1,0.1)\;s\\
Speed Regulator                     & (1.5,0.5) Hz/pu\\
Diesel Generator power rate ($\gamma_2$)& 0.5\;pu\\
Energy storage  power rate ($\gamma_3,\gamma_4$)& (0.2,0.15)\; pu\\
External input $(\Delta P_L-\Delta P_{RES})$ power rate $\gamma_\omega$ & 0.2\;pu.
\end{tabular}

\end{table}

Three different types of attacks are considered: 
\begin{itemize}
    \item[1] \emph{Random AGC setpoint attack:} an adversary is able to replace all AGC setpoints sent to the generation and energy storage units with random signals drawn from an uniform distribution.
    \item[2] \emph{Optimal AGC setpoint attack:} an adversary replaces all AGC setpoints with an optimal sequence carefully designed to maximize the frequency deviation after 50 iterations (100 seconds). The attack sequence is found by solving the following optimization problem:
    \begin{align*} \max_{u^a(1),\ldots,u^a(50)}&{x_1(50)}\\
    s.t. \;\;x(k+1)&=Ax(k)+Bu^a(k)+H\omega(k)\\
         |u_i^a(k)|&\leq \gamma_i,\;\forall i=1,\ldots, 4, k=1,\ldots,50
    \end{align*}
    
    \item[3] \emph{Optimal sensor attack:} the adversary injects $\delta^a(k)$ to the frequency measurements used to compute the AGC setpoints. The attack sequence is designed by solving the following optimization problem:
    \begin{align*} \max_{\delta^a(1),\ldots,\delta^a(50)}&{x_1(50)}\\
    s.t.\;\; x(k+1)&=Ax(k)+Bu(k)+H\omega(k)\\
         ACE(k)&=\mathcal{B}(x_{1}(k)+\delta^a(k))\\
         AGC(k)&=K_PACE(k)+K_I\sum\nolimits_{0}^{50}ACE(k)\\ 
         u_i(k)&=\beta_iAGC(k)\\
         |u_i(k)|&\leq \gamma_i,\;\forall i=1,\ldots, 4, k=1,\ldots,50.
    \end{align*}
\end{itemize}
 
 \subsection{Normal Operation}
\begin{figure}[h]
     \centering
     \includegraphics[width=\columnwidth]{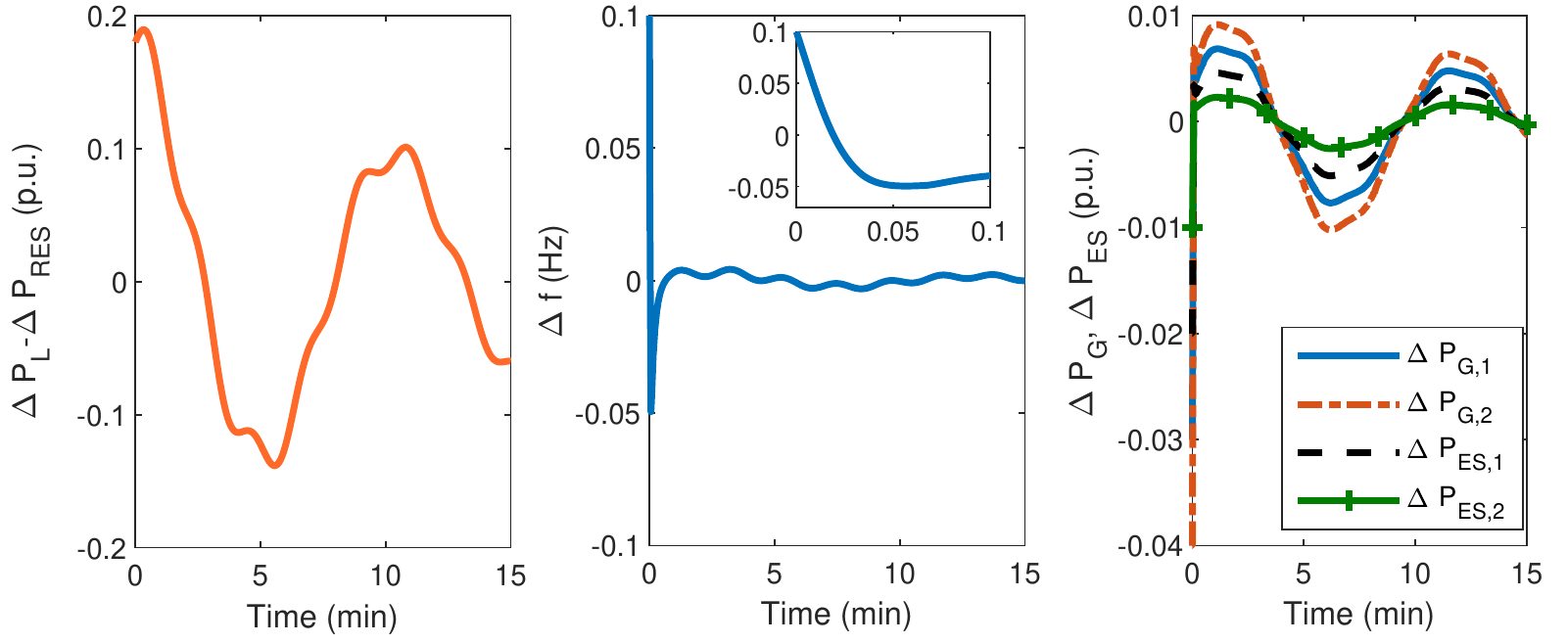}
     \caption{Normal operation of the power grid with frequency control.  }
     \label{fig:normal}
 \end{figure}

The optimal resilient constraints obtained by solving the optimization problem in \eqref{thmdangerous} are $ \hat \gamma=[0.1,0.38,0.2,0.15]$. Figure \ref{fig:normal} illustrates the frequency deviation,  unknown external input $\Delta P_{L}-\Delta P_{RES}$, and the AGC response during 15 min with the obtained resilient constraints. Notice that the AGC is able to bring frequency deviations close to zero and the controller actions are never saturated.  
 
\subsection{Attack Impact without Resilient Constraints} 
\begin{figure}[h]
     \centering
     \includegraphics[width=1\columnwidth]{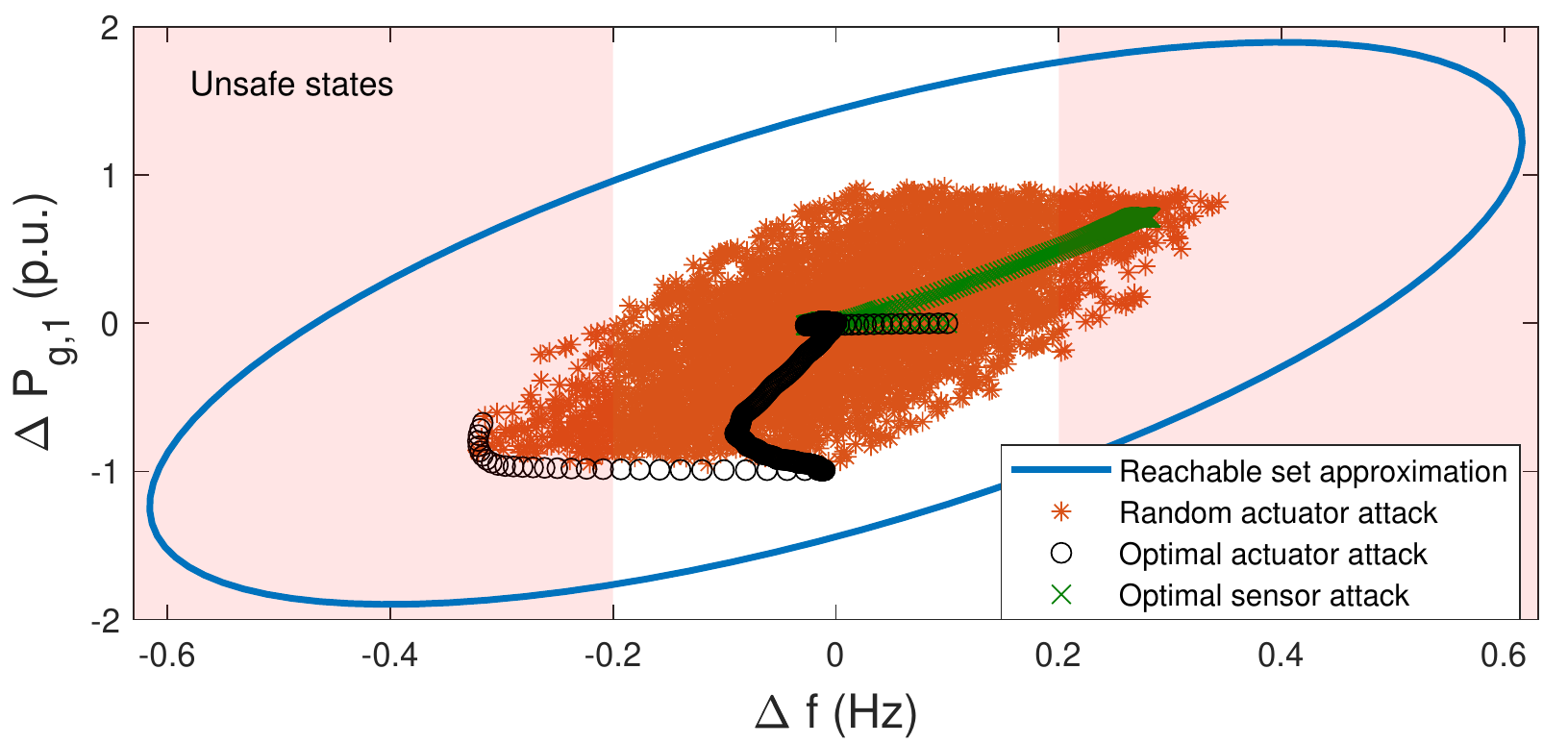}
     \caption{Reachable set without the proposed resilient constraints. Unsafe states can be attained by multiple attack sequences. }
     \label{fig:reachable_no_defense}
 \end{figure}

\begin{figure}[h]
     \centering
     \includegraphics[width=\columnwidth]{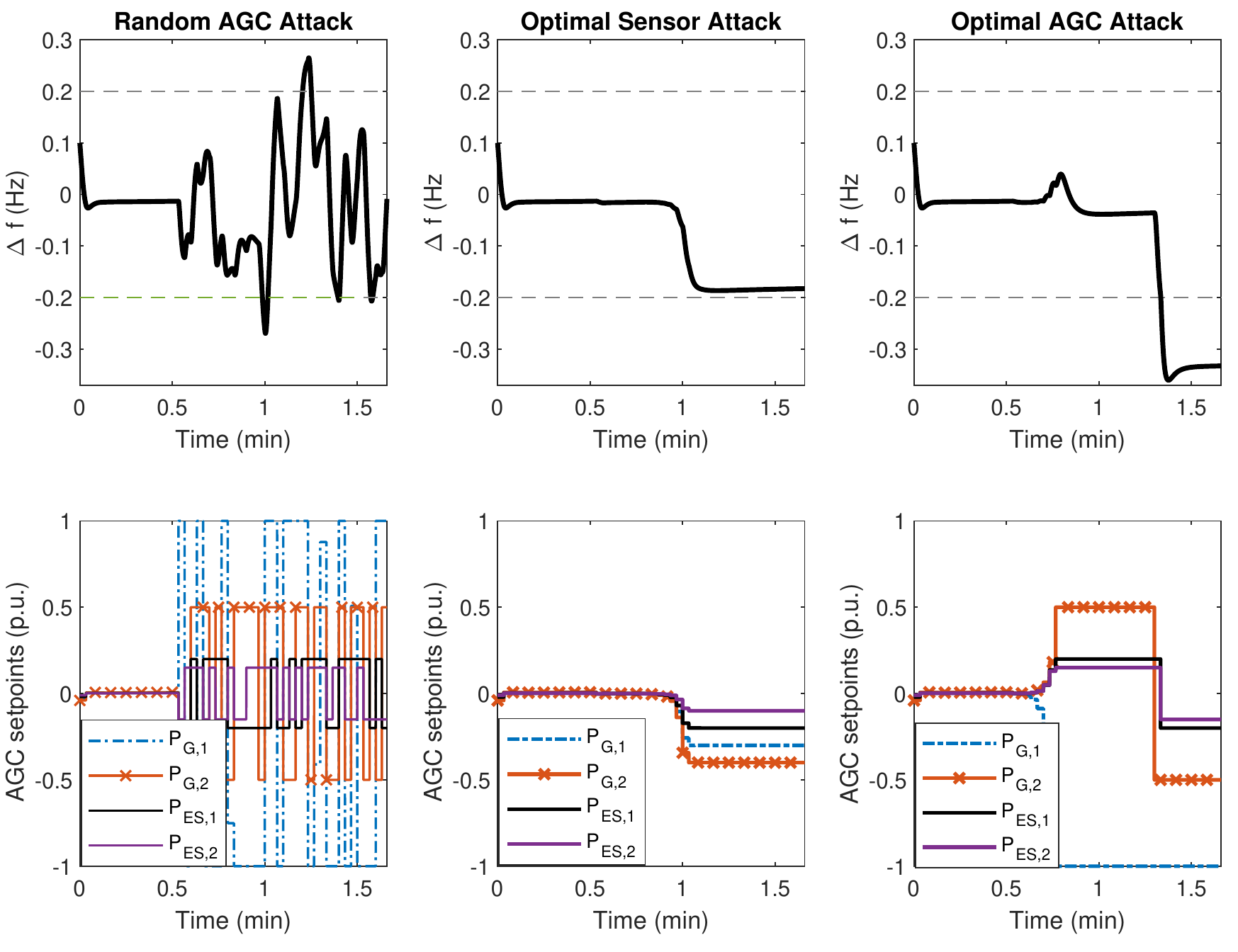}
     \caption{Frequency deviation and AGC setpoints for different attack scenarios.}
     \label{fig:attack_no_defense}
 \end{figure}

 In order to illustrate the potential impact caused by malicious attackers, let us consider the case without the resilient constraints. 
 Figure \ref{fig:reachable_no_defense} depicts the projection of the ellipsoidal reachable set approximation with respect to frequency deviation $\Delta f$ and $\Delta P_{g,1}$ with the original physical constraints. The set of unsafe constraints corresponds to $\Delta f\geq 0.2\;Hz$, $\Delta f\leq -0.2\;Hz$. The state deviation caused by the three attack scenarios is also depicted in Fig. \ref{fig:reachable_no_defense}, where $1.000$ iterations of the random attack were performed.  Notice that the reachable set overlaps  the set of unsafe states such that there exist multiple attack sequences that can drive frequency deviations to dangerous values. 
 Figure \ref{fig:attack_no_defense} illustrates the frequency deviation and AGC setpoints for the three attack scenarios. In particular, the random AGC attack and the optimal AGC attack can cause frequency deviation below $-0.3$ Hz, which surpasses the safety limits. 
 
 \subsection{Attack Impact With Resilient Constraints}

\begin{figure}[h]
     \centering
     \includegraphics[width=1\columnwidth]{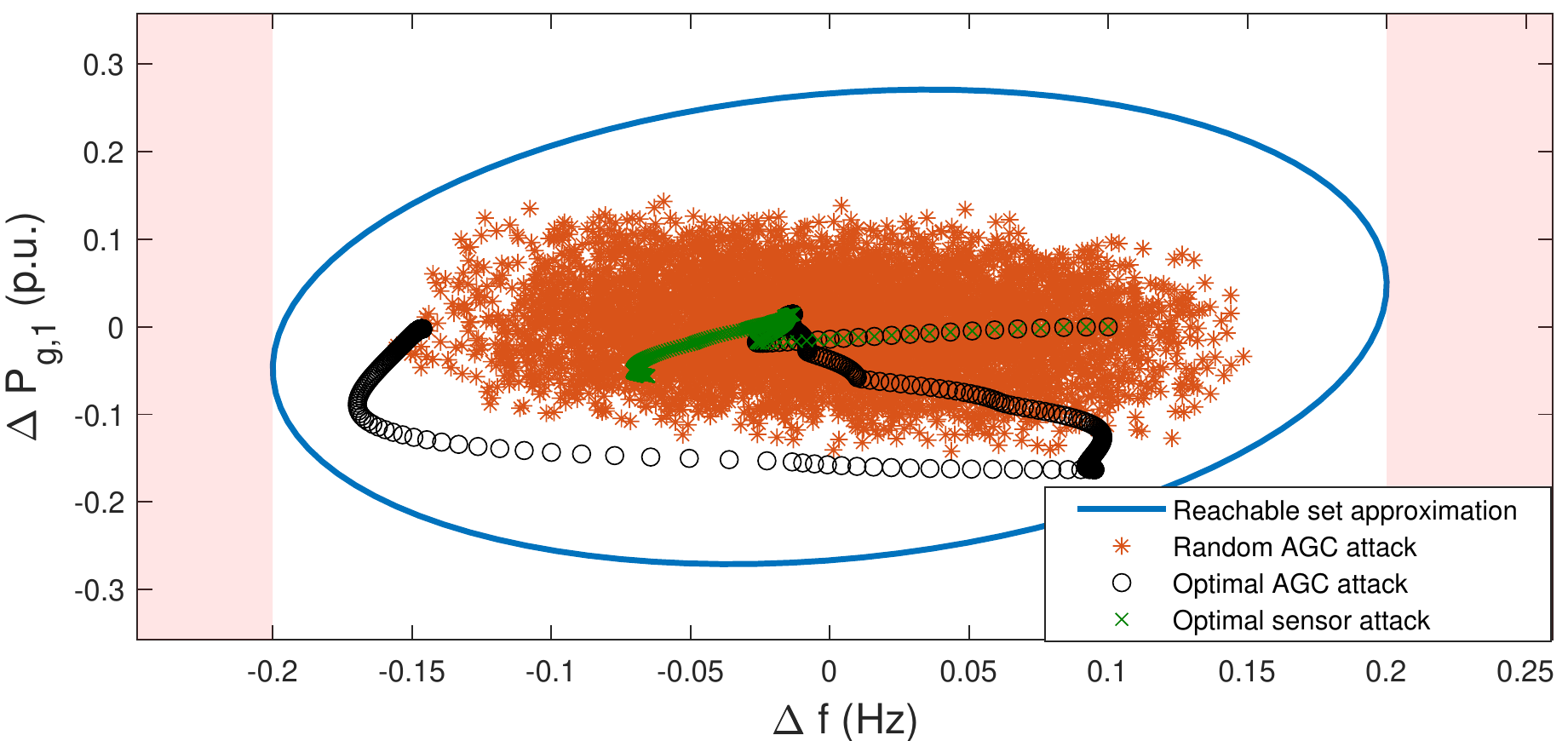}
     \caption{Reachable set with the proposed resilient operating constraints. No attack can drive the system to unsafe states.}
     \label{fig:reachable_defense}
 \end{figure}

\begin{figure}[h]
     \centering
     \includegraphics[width=\columnwidth]{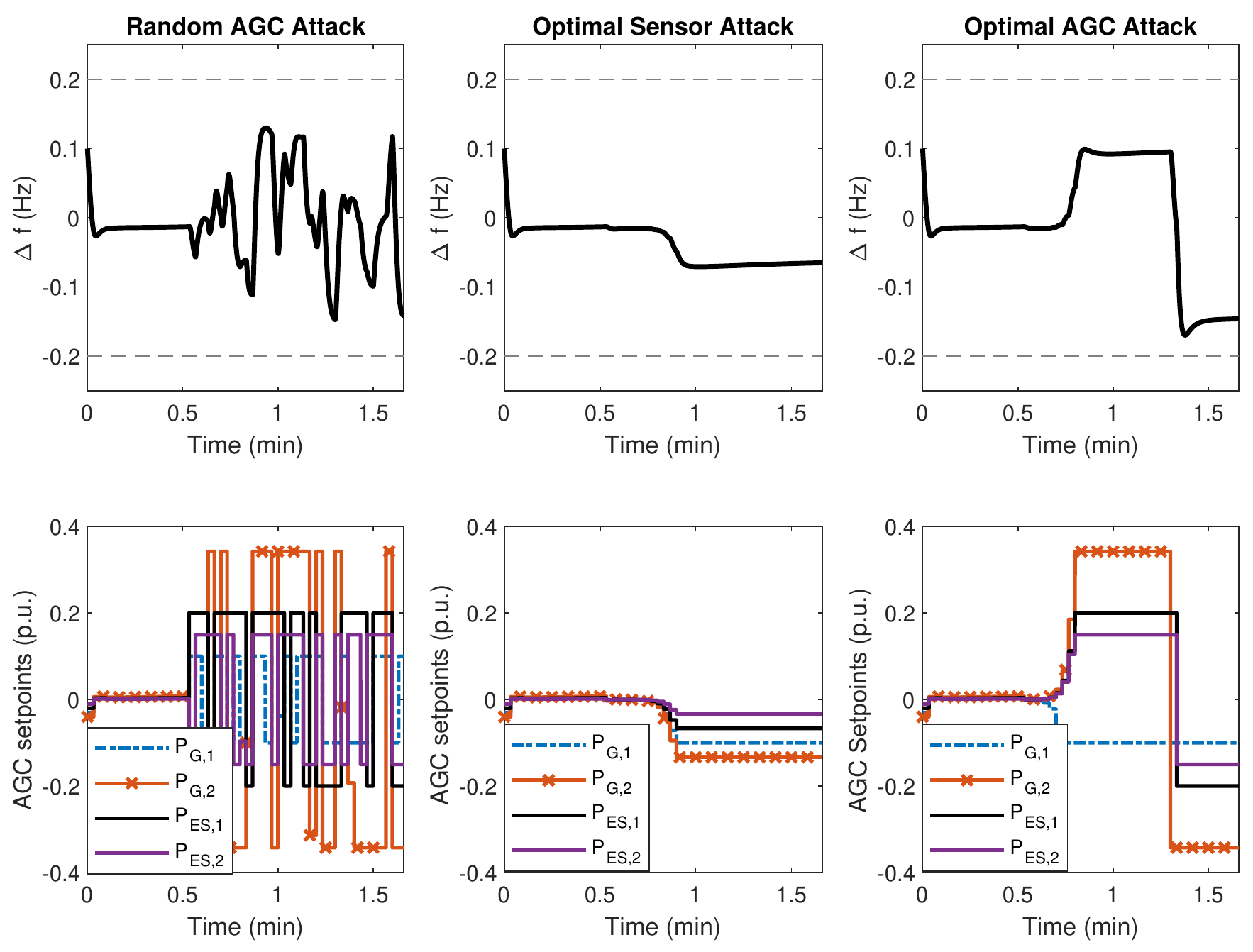}
     \caption{Frequency deviation and AGC setpoints with the proposed resilient operating constraints for different attack scenarios. }
     \label{fig:attack_defense}
 \end{figure}

The resilient reachable set is illustrated in Fig. \ref{fig:reachable_defense} with the three attacks introduced above. Notice that the ellipsoidal approximation does not overlap the unsafe states guaranteeing that the frequency deviation remains within desired limits. As a consequence, there does not exist any attack sequence that can drive the system states to unsafe states. In fact, Fig. \ref{fig:attack_defense} illustrates that the optimal AGC attack can only cause a frequency deviation up to $-0.18$ Hz and it requires a very specific attack sequence. On the other hand, the optimal sensor attack can only cause a maximum deviation of -0.08 Hz. 
 
 \section{Conclusions}\label{sec:conclusions}
This paper presented a methodology that allows to  compute resilient operating constraints for generation units and energy storage systems that guarantee that no attack affecting the AGC setpoints directly or indirectly can drive the system  to unsafe states of operation. Typically, the computation of these types of constraints becomes an NP-hard combinatorial problem when the range of operation  is continuous, forming a bounded but infinite set.  However, the proposed methodology uses ellipsoidal approximations to solve a convex optimization problem with LMI constraints that find the group of resilient operating constraints based on the dynamic description of the system and a defined set of unsafe states. The simulation results have shown that without resilient constraints it is possible for an attacker to design an attack sequence that can drive the frequency deviation below $0.2\;Hz$. However, by imposing the resilient constraints found with the proposed methodology, the resilient operation of the power system with AGC control is guaranteed and the frequency deviation remains within desired limits. In fact, there does not exist any attack sequence that can force frequency deviations to violate safety conditions. In addition, the obtained constraints do not degrade the normal operation of the power system.

\bibliographystyle{IEEEtran}

\appendices
\section{Proof Theorem 1}\label{app:thm1}
In order to prove Theorem 1,  it is necessary to restate the following lemma adapted from  \cite{that2013reachable}.

\begin{lemma}\label{lemm:Vk}
Let $V(k)$ be  a  positive  definite  function, for $k\in\mathbb{N} $ where $V(0)=  0$. Let $|u_{i,k}|\leq  \gamma_i$ for $i=1,\ldots,m$ denote some bounded inputs.  If  there  exists  a  constant $a\in(0,1)$ such that the following holds,
\vspace{-5pt}
\[V({k+1})-aV(k)-\frac{(1-a)}{m}\sum\limits_{j=1}^m{\frac{u_{j}(k)^2}{\gamma_i^2}}\leq 0,
\vspace{-5pt}\]
then $V(k)\leq 1$ for all $k$.
\end{lemma}
\vspace{5pt}

Suppose that we define the function $V(k)=x(k)^\top W^{-1} x(k)$. Thus, from  Lemma \ref{lemm:Vk} and replacing with \eqref{eq:x}, we have that, if
\begin{align*}x(k)^\top A^{\top}W^{-1}Ax(k)+2x(k)^\top A^{\top}W^{-1}B u(k)+\ldots\\
 +2x(k)^\top A^\top W^{-1} H \omega(k)+2u(k)^\top B^\top W^{-1}H\omega(k)+\ldots \\
 u(k)^\top B^\top W^{-1}B u(k)+\omega(k)^\top H^\top W^{-1}H\omega(k)-\ldots\\
 -\frac{(1-a)}{m}\left(u(k)R_u^{-1}u(k)-\omega(k)^\top(\gamma_\omega^2)^{-1}\omega(k)\right)-\ldots\\
 -ax(k)^{\top}W^{-1}x(k)\leq 0
\end{align*}
then $x(k)^\top W^{-1}x(k)\leq 1$.
Let $G=[B,H]$, $\hat R=diag(\hat \gamma_1,\ldots,\hat \gamma_m,\gamma_{\omega})$, and $\widetilde u(k)=[u(k)^\top,\omega(k)]^\top$. Therefore, the inequality constraint can be rewritten in matrix form as follows:
\begin{equation}\label{eq:J}
[x,\widetilde u]\underbrace{\begin{bmatrix}
aW^{-1}-A^\top W^{-1}A & -A^\top W^{-1}G \\
-G^\top W^{-1} A & \frac{(1-a)}{m}\hat R^{-1}-G^\top W^{-1}G & \end{bmatrix}}_{\mathcal{J}}\begin{bmatrix}x\\\widetilde u\end{bmatrix}\geq 0
\end{equation}
Notice that the condition in \eqref{eq:J} is satisfied if $\mathcal{J}>0$. In order to remove the inverse terms, it is possible to use a mathematical tool named the Schur complement, which allows to find an equivalent matrix $\ \mathcal{\hat J}$, such that if $\mathcal{\hat{J}}\geq0$, it implies that  $\mathcal{J}\geq 0$. 

Calculating the Schur complement to \eqref{eq:J} leads to:
\[\mathcal{\hat J}=\begin{bmatrix}
   aW^{-1}& 0& A^\top W^{-1} \\
   0&    \frac{(1-a)}{m}R^{-1} & G^\top W^{-1}\\
   W^{-1}A & W^{-1}G & W^{-1}
\end{bmatrix}\geq 0
\]
Finally, applying the congruence transformation $\mathcal{\hat J}\rightarrow \mathcal{T}^\top \mathcal{\hat J} \mathcal{T}$, where \[\mathcal{T}=\begin{bmatrix}W&0&0\\0&R&0\\0 &0&W\end{bmatrix},\]
results in the first condition  in \eqref{thmdangerous}. Therefore, if there exists a positive definite matrix $W$, the reachable set is contained inside the ellipsoid $x^\top W^{-1} x\leq 1$.
The second constraint comes from the distance $d_{\mathcal{E},i}$ defined in \eqref{eq:distance}. 
Therefore,  $d_{\mathcal{E},i}\geq 0$ is equivalent to $|g_i|-\sqrt{c_i^\top W c_i}\geq0$, which leads to 
\[c_i^\top Wc_i\leq g_i^2, \]
which must hold for all the hyperplanes $i=1,\ldots,\kappa$. 

\hfill $\blacksquare$

\end{document}